# Superluminal k-gap solitons

# in nonlinear photonic time-crystals

Yiming Pan†[1], Moshe-Ishay Cohen†[1], Mordechai Segev[1,2]

1. *Physics Department and Solid State Institute, Technion, Haifa 32000, Israel*
2. *Department of Electrical and Computer Engineering, Technion, Haifa 32000, Israel*

**Abstract**

We propose superluminal solitons residing in the momentum gap (k-gap) of nonlinear photonic time-crystals. These gap solitons are structured as plane-waves in space while being periodically self-reconstructing wavepackets in time. The solitons emerge from modes with infinite group velocity causing superluminal evolution, which is opposite to the stationary nature of the analogous Bragg gap soliton residing at the edge of an energy gap (or a spatial gap) with zero group velocity. We explore the faster-than-light pulsed propagation of these k-gap solitons in view of Einstein's causality by introducing a truncated input seed as a precursor of signal velocity forerunner, and find that the superluminal propagation of k-gap solitons does not break causality.

Photonic time-crystals (PTCs) [1–11] are dielectric media whose permittivity ($\epsilon$) is modulated periodically in time, causing time-reflected and time-refracted waves [3] to interfere, giving rise to Floquet modes associated with the momentum bands and bandgaps (also called k-gaps). Naively, PTCs seem similar to one-dimensional spatial photonic crystals (SPCs), whose dispersion relations are determined by periodic variation of the dielectric permittivity in space (Fig. 1a), with energy (frequency) gap where the associated Bloch modes have complex momentum rendering them localized in space. This analogy is misleading, as SPCs and PTCs differ in several fundamental aspects. First, waves propagating in a dielectric SPC conserves energy (because the medium is stationary in time) but exchange momentum with the spatial lattice, whereas PTCs do not conserve energy (the modulation breaks time-translation symmetry) but conserve momentum. Second, the amplitudes of the Bloch modes in the frequency gap of a SPC always decay in space with the distance from the source, whereas the Floquet modes in the momentum k-gap of a PTC (Fig. 1b) exhibit exponential growth (or decay) with increasing time from the moment of excitation. In fact, the gap modes of PTCs draw energy from the modulation (or lose their energy to it). The unavoidable presence of exponentially growing (and decaying) gap modes in PTCs is intimately related to causality (see intuitive explanation in [12]). In the past few years, PTCs have been drawing growing research interest [10–21], and the recent experimental progress with very large permittivity changes at few-femtosecond time scales in epsilon-near-zero materials [22–28] suggests that PTCs at optical frequencies will be observed in the near future.

The presence of a momentum gap (k-gap) in PTCs suggests the existence of gap solitons in PTCs, in analogy to SPCs. Gap solitons are self-trapped entities known to reside in the bandgap of nonlinear periodic systems such as fiber gratings [29–34] and waveguide arrays [35–41] (Fig. 1c). Gap solitons of SPCs inherit some of their features from the Bloch modes associated with the band edge in the linear system; most profoundly, gap solitons in SPCs are always stationary with zero group-velocity. Is it possible to have gap solitons in the momentum gap of a nonlinear time-varying photonic media? And if such k-gap solitons do exist in PTCs, will they be stationary as gap solitons in SPCs, or will their group velocity be infinite, inherited from the Floquet modes of the band edge of PTCs? Finally, if their group velocity is infinite, how can they be reconciled with causality?

Here, we find gap solitons in the momentum gap of nonlinear photonic time-crystals. In spatially-homogeneous PTCs, these k-gap solitons are structured as finite wavepackets in time but plane waves in space. We find that the k-gap solitons are superluminal: their peaks travel at speeds that are faster than light. For a broad input beam, the group velocity of the k-gap solitons diverges, becoming infinite for a plane-wave input. The superluminal propagation of the k-gap soliton seems to contradict Einstein's special relativity, and raises fundamental questions on the physical essence of such solitons and of what can be unraveled by experiments. We reconcile our findings with special relativity through Sommerfeld's forerunners, and suggest interesting experiments for launching pulsed beams whose peak intensity propagates faster than light.

We begin from Maxwell's equations in a nonlinear medium with an instantaneous nonlinearity, where the linear part of the refractive index is modulated periodically in time. For simplicity, we assume that the nonlinearity is the optical Kerr effect, but the ideas described henceforth are applicable to any local nonlinearity. In one-dimensional media, this yields the relation between the electric displacement vector $D$ and the electric field of the light $E$, as $\vec{D} = \epsilon(t, |E|^2)\vec{E} = \epsilon_0(\epsilon_1(t) + \chi^3 \epsilon_0 |E|^2)\vec{E}$, with $\chi^3$ being the Kerr coefficient. Here, $\epsilon_1$ is the linear dielectric constant, chosen to be spatially homogeneous but periodically modulated in time $\epsilon_1(t) = \epsilon_r(1 + \delta_1 \cos \Omega t)$, in which $\epsilon_r$ is the mean value of the permittivity, $\delta_1 < 1$ a small real dimensionless quantity and $\Omega = 2\pi/T$ the modulation frequency, and $T$ the modulation period. For simplicity, we assume that the medium is isotropic, and henceforth drop the vector signs. We expand $E$ in terms of $D$ as

$$E = \frac{D}{\epsilon_0 \epsilon_1} - \frac{\Gamma_3 D^3}{\epsilon_0} = \frac{D}{\epsilon_0 \epsilon_1} - \frac{\chi^3 D^3}{\epsilon_0^2 \epsilon_1^4} \tag{1}$$

where we define the nonlinear coefficient as $\Gamma_3 = \frac{\chi^3}{\epsilon_0 \epsilon_1^4}$. From Maxwell equations, we obtain:

$$\frac{\partial^2 D}{\partial t^2} = \frac{1}{\mu_0} \frac{\partial^2}{\partial x^2} \left( \frac{D}{\epsilon_0 \epsilon_1} - \frac{\chi_3 D^3}{\epsilon_0^2 \epsilon_1^4} \right) \tag{2}$$

This form is different from the conventional equation $\frac{\partial^2}{\partial t^2}(\epsilon_0 \epsilon_1(t) E) = \frac{1}{\mu_0} \frac{\partial^2}{\partial x^2} E$ in terms of $E$ [11]. The reason to prefer $D$ over $E$ is that, for PTC, $D$ is continuous whereas $E$ is not necessarily continuous. We note that the modulation parameters $\delta_1$ is small, so the influence of periodic modulations on the Kerr term is negligible, as long as the Kerr coefficient $\chi^3$ is small. Therefore, we simplify Eq. 2 to

$$\frac{1}{c^2} \frac{\partial^2 D}{\partial t^2} = (1 - \delta_1 \cos \Omega t) \frac{\partial^2 D}{\partial x^2} - \beta |D|^2 \frac{\partial^2 D}{\partial x^2} \tag{3}$$

with the speed of light in the medium $c = 1/\sqrt{\mu_0 \epsilon_0 \epsilon_r} = c_0/\sqrt{\epsilon_r} = c_0/n_0$. We approximate $\epsilon_1^{-1} = (1 - \delta_1 \cos \Omega t)/\epsilon_r$ and, in the last term only consider the contribution of the (automatically phase-matched) self-phase-modulation term and redefine the corresponding nonlinear coefficient $\beta = 3\chi^3/\epsilon_0 \epsilon_r^3$.

To obtain the k-gap soliton, we first find the dispersion relation from (3), and subsequently derive an effective nonlinear Schrödinger-like equation within, or close to, the photonic k-gap. There are many alternative treatments [31–33,42,43]; we use the nonlinear coupled-mode theory with a slowly varying envelope approach in terms of seeking solutions for the dielectric field as the sum of suitably modulated forward and backward waves,

$$D(x,t) = A_f(x,t)e^{ik_0 x - i\Omega t/2} + A_b(x,t)e^{ik_0 x + i\Omega t/2} + c.c., \qquad (4)$$

with $k_0 = \Omega/2c$. Substituting this ansatz into (3) and applying the slowly varying envelope approximation, we find that $A_{f,b}$ obey the following pair of coupled-mode equations:

$$\begin{aligned}
+i\left(\frac{1}{c}\frac{\partial A_f}{\partial t} + \frac{\partial A_f}{\partial x}\right) + \kappa A_b + \gamma\left(|A_f|^2 + 2|A_b|^2\right)A_f = 0 \\
-i\left(\frac{1}{c}\frac{\partial A_b}{\partial t} - \frac{\partial A_b}{\partial x}\right) + \kappa A_f + \gamma\left(|A_b|^2 + 2|A_f|^2\right)A_b = 0
\end{aligned} \qquad (5)$$

With the coupling coefficient $\kappa = \delta_1 \Omega/8c$ and the nonlinear coefficient $\gamma = \beta\Omega/4c$ incorporating the Kerr nonlinear term. We note that, if the nonlinear term has the same magnitude for the self-phase and cross-phase modulation, this equation is fully integrable with the known solutions of Manakov solitons [44–49].

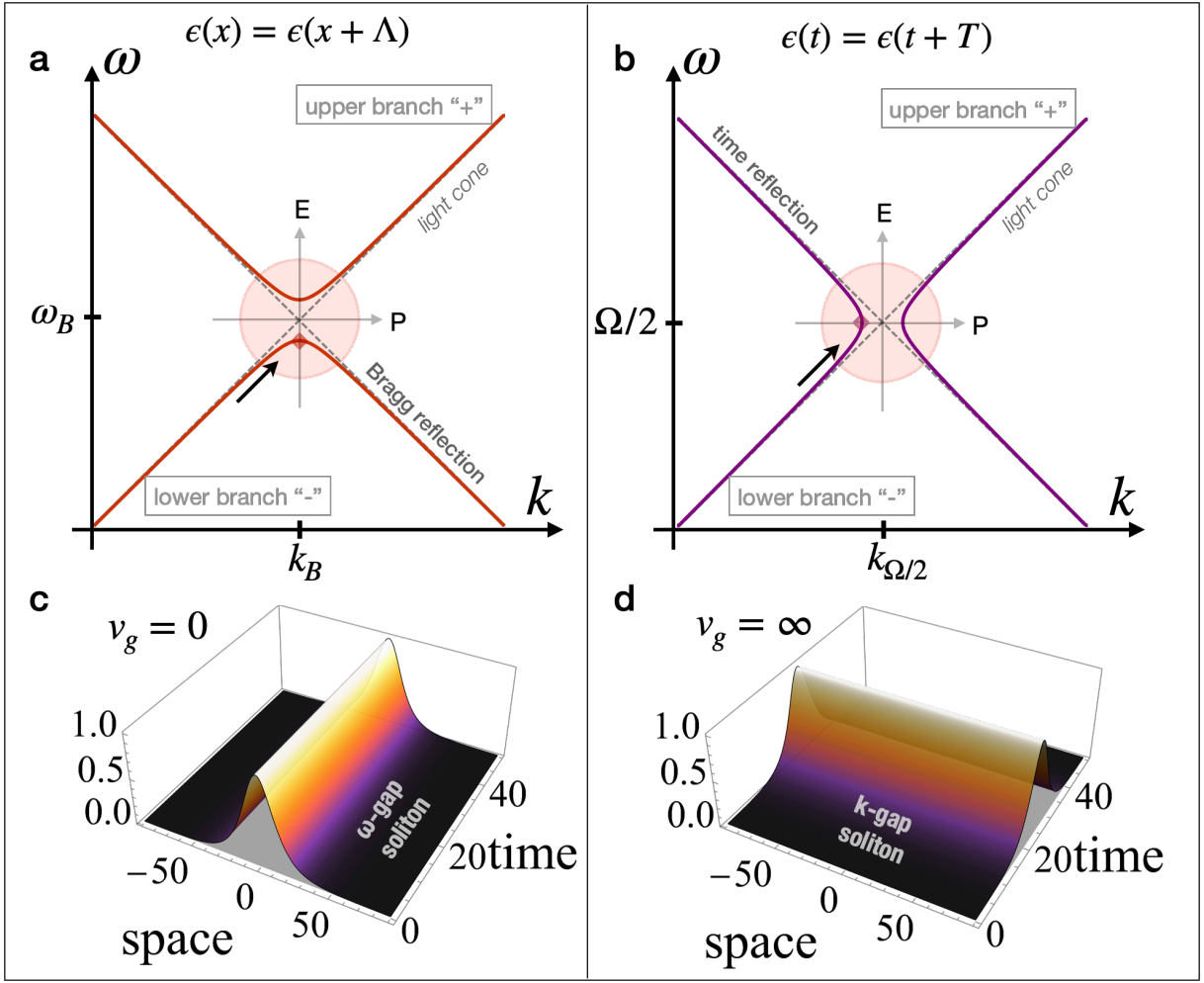

Fig. 1: (a) Band structure of a PC created by spatial periodic modulation $\epsilon(x) = \epsilon(x + \Lambda)$. (b) Band structure of a PTC created by the temporal periodic modulation $\epsilon(t) = \epsilon(t + T)$. (c) Self-focusing Kerr nonlinearity gives rise to a Bragg ($\omega$-gap) soliton emerging at the band edge where the group velocity ($v_g$) is zero. (d) Self-focusing Kerr nonlinearity gives rise to a k-gap soliton emerging at the band edge where the group velocity is infinite.

**Locating the k-gap** - To find the photonic band with k-gap, we define the two components field $\psi$ as $\psi = (A_f, A_b)^T$. Seeking solutions for the linear part of (5), of the shape $\psi = \chi e^{i(Px - Et)}$ with $\chi$ being a spinor, we insert this ansatz into (5) in the case $\gamma = 0$, and obtain the energy-momentum dispersion relation,

$$P^{(\pm)} = \pm\sqrt{\left(\frac{E}{c}\right)^2 + \kappa^2} \tag{6}$$

As expected from [11], there is a momentum gap ("k-gap") for $|P| < \kappa$. The k-gapped photonic band is plotted in Fig. 1c, where the sign $\pm$ denotes the upper and lower branches, respectively. Let us take a look at the local group velocity and the group velocity dispersion (GVD) around the band edge of the k-gap. We choose a point $\left(E_0 + \frac{\Omega}{2}, P_0 + k_0\right)$ on the band, expand the dispersion (6) around this point to the second-order, and obtain

$$\begin{aligned}v_g^{(\pm)} &= \left(\frac{\partial P^{(\pm)}}{\partial E}\right)^{-1} = \pm\left(\frac{\sqrt{(E_0/c)^2+\kappa^2}}{E_0/c}\right)c, \\ \text{GVD}^{(\pm)} &= \left(\frac{\partial^2 P^{(\pm)}}{\partial E^2}\right) = \pm\frac{1}{c^2\sqrt{\left(\frac{E_0}{c}\right)^2+\kappa^2}}.\end{aligned} \tag{7}$$

Two points must be addressed here. First, we notice the group velocity $v_g^{(\pm)}$ becomes infinitely large as $E_0$ goes to 0 (i.e., when the optical frequency goes to half of the modulation frequency $\omega \to \Omega/2$). The infinite group velocity indicates that any physical solution around the bandgap must be a moving solution. In this sense, the energy carried by the soliton should (presumably) travel at a speed faster than the speed of light, as shown in Fig. 1d. This poses an intriguing question: the propagation velocities of gap solitons arise from those of the corresponding linear modes of the system, hence would the k-gap solitons be superluminal, and if so, how can this be reconciled with special relativity? Notably, the Bragg gap solitons in 1D photonic crystals are stationary, because the group velocity of the Bloch mode at the band edge $v_g \to 0$ (Fig. 1a). Indeed, as we show later on, the k-gap solitons in PTCs are superluminal (Fig. 1b). The second point is that the group velocity dispersion (GVD) is nonzero at the band edges, that is, $\text{GVD}^{(\pm)} = \pm 1/\kappa c^2 = \pm 8/\delta_1 \Omega c$. Thus, even though the modes around the band edge are superluminal they still experience dispersive effects. Namely, even though the first derivative of the dispersion relation is infinite, the second derivative still plays an important role, and as we show below – its presence is crucial to the formation of k-gap solitons. The sign of the GVD denotes normal or anomalous dispersion, respectively. Correspondingly, the spinor eigenvectors $\chi^{(\pm)}$ associated with $P_\pm$ are,

$$\chi^{(+)} = \frac{1}{\sqrt{2\sqrt{\frac{E^2}{c^2}+\kappa^2}}} \begin{pmatrix} \frac{\kappa}{\sqrt{\sqrt{\frac{E^2}{c^2}+\kappa^2}-\left(\frac{E}{c}\right)}} \\ -\sqrt{\sqrt{\frac{E^2}{c^2}+\kappa^2}-\left(\frac{E}{c}\right)} \end{pmatrix} \equiv \begin{pmatrix} \chi_1^{(+)} \\ \chi_2^{(+)} \end{pmatrix}$$

$$\chi^{(-)} = \frac{1}{\sqrt{2\sqrt{\frac{E^2}{c^2}+\kappa^2}}} \begin{pmatrix} \sqrt{\sqrt{\frac{E^2}{c^2}+\kappa^2}-\left(\frac{E}{c}\right)} \\ \frac{\kappa}{\sqrt{\sqrt{\frac{E^2}{c^2}+\kappa^2}-\left(\frac{E}{c}\right)}} \end{pmatrix} \equiv \begin{pmatrix} \chi_1^{(-)} \\ \chi_2^{(-)} \end{pmatrix}$$

(8)

Note the normalization $\chi^{(+)}\chi^{(+)} = \chi^{(-)}\chi^{(-)} = 1$ and the orthogonality $\chi^{(+)}\chi^{(-)} = 0$. The solutions at the band edges are of the form $\chi^{(+)} = (1,-1)^T/\sqrt{2}$ for the upper branch, and $\chi^{(-)} = (1,1)^T/\sqrt{2}$ for the lower branch. These solutions can be described as a uniform standing wave: the wave does not move at all, and all points in space have the same amplitude at a given time, but the amplitude oscillates in time at half the modulation frequency with $D^{(-)} = e^{iP^{(-)}x}\cos\left(\frac{\Omega t}{2}\right)$, and $D^{(+)} = e^{iP^{(+)}x}\sin\left(\frac{\Omega t}{2}\right)$. Henceforth we present the lower-branch k-gap solitons, while the upper branch solitons are discussed in the Supplementary Material.

**Solitons in the k-gap.** Next, we derive the nonlinear Schrödinger equations (NLSE) from the nonlinear coupled-mode equations (5). For the lower band branch (−), we search for a solution of the form

$$\psi = a(x,t)\chi^{(-)}e^{i(-P_0 x - E_0 t)} \tag{9}$$

with $P_0 = \sqrt{\left(\frac{E_0}{c}\right)^2 + \kappa^2}$. Substituting (9, 8) into (5), we obtain the NLSE for the slowly varying amplitude $a(x,t)$,

$$\left(-i\frac{\partial}{\partial x} - \frac{i}{v_g}\frac{\partial}{\partial t} - \frac{\text{GVD}}{2}\frac{\partial^2}{\partial t^2}\right)a + \alpha|a|^2 a = 0 \tag{10}$$

where the nonlinear coefficient $\alpha = \gamma\left(3 - \frac{E_0/c}{E_0^2/c^2 + \kappa^2}\right)/2$. The detailed derivations are given in the Supplementary Material [X]. To solve the NLSE (10) explicitly, we make the substitution $T = t - \frac{x}{v_g}$, $X = x$, and thus, the nonlinear equation becomes a standard form $\left(-i\frac{\partial}{\partial X} - \frac{\text{GVD}}{2}\frac{\partial^2}{\partial T^2}\right)a + \alpha|a|^2 a = 0$. We notice that in (7) the GVD $< 0$ for the lower branch (−), so that to construct a bright soliton solution, we need a focusing Kerr nonlinear $\alpha > 0$ (or $\gamma > 0$) [36,38,40]. Otherwise, we would obtain the dark k-gap solitons. The bright soliton solution for (10) is

$$a = u_0 \text{sech}\left(\frac{T}{\tau_0}\right)e^{iqX} = u_0 \text{sech}\left(\frac{t - x/v_g}{\tau_0}\right)e^{iqx} \tag{11}$$

with the relations $\tau_0 = \tau_0(u_0) = \frac{1}{u_0}\sqrt{\frac{|\text{GVD}|}{\alpha}}$, $q = q(u_0) = \frac{\alpha u_0^2}{2}$. Thus, we obtain the soliton spinor wavefunction (9) $\psi = u_0 \text{sech}\left(\frac{t - x/v_g}{\tau_0}\right)\chi^{(-)}e^{i(-P_0 + q)x}e^{-iE_0 t}$, and the corresponding electric displacement vector (4),

$$D(x,t) = A_f(x,t)e^{ik_0 x - i\Omega t/2} + A_b(x,t)e^{ik_0 x + i\Omega t/2} + c.c.$$
$$= u_0 \operatorname{sech}\left(\frac{t - x/v_g}{\tau_0}\right)\left(\chi_1^{(-)} e^{i(k_0 - P_0 + q)x} e^{-i\left(\frac{\Omega}{2} + E_0\right)t} + \chi_2^{(-)} e^{i(k_0 - P_0 + q)x} e^{i\left(\frac{\Omega}{2} - E_0\right)t} + c.c.\right) \quad (12)$$

At the band edge ($E_0 \to 0$), the parameters are: $1/v_g = 0$, $|\text{GVD}| = \frac{1}{\kappa c^2}$, $\alpha = \frac{3\gamma}{2}$, $\tau_0 = \frac{1}{cu_0}\sqrt{\frac{2}{3\gamma\kappa}}$, $q = \frac{3\gamma u_0^2}{4}$ and $\chi_1^{(-)} = \frac{1}{\sqrt{2}}$, $\chi_2^{(-)} = \frac{1}{\sqrt{2}}$. So, the bright k-gap soliton of the lower branch is

$$D(x,t) = \frac{u_0}{\sqrt{2}} \operatorname{sech}\left(\frac{t}{\tau_0}\right)\left(e^{i(k_0 + \kappa - q)x} e^{-i\left(\frac{\Omega}{2}\right)t} + e^{i(k_0 + \kappa - q)x} e^{i\left(\frac{\Omega}{2}\right)t} + c.c.\right)$$
$$= \frac{2u_0}{\sqrt{2}} \operatorname{sech}\left(\frac{t}{\tau_0}\right)\left(\cos\left(\left(k_0 - \kappa + \frac{3\gamma u_0^2}{4}\right)x - \frac{\Omega t}{2}\right) + \cos\left(\left(k_0 - \kappa + \frac{3\gamma u_0^2}{4}\right)x + \frac{\Omega t}{2}\right)\right) \quad (13)$$

where $u_0$ is the soliton peak amplitude. We find many interesting effects from (13). First, the k-gap soliton at the band edge has the temporal form $\operatorname{sech}(t/\tau_0)$, as shown in Fig. 1d, but has no spatial dependence, i.e., the field is uniform in space. We notice that our k-gap soliton is fundamentally different from the stationary Bragg solitons residing in the $\omega$-gap [29–34] (Fig. 1b), intrinsically due to the infinite group velocity at the band edge of the k-gap. Second, the soliton consists of two counterpropagating pulses that are simultaneously generated. The underlying physics of pair generation stems from the momentum conservation in PTCs [15,50]. Third, we see that the full electromagnetic field of soliton is oscillating at frequency and wavevector:

$$\omega = \frac{\Omega}{2}, \quad k = k_0 - \kappa + q = k_0 - \kappa + \frac{3|\gamma|u_0^2}{4} \in k\text{-gap} \quad (14)$$

The k-gap soliton shows a subharmonic response to the time modulation, with a frequency locked at $\Omega/2$, but its wavenumber can be anywhere in the "linear" gap, depending on the magnitude and the sign of the nonlinearity! ($\gamma > 0$ in our case). Fourth, notice that the intensity of the soliton goes down monotonically in time after it reaches its peak. This is very surprising, as in linear PTC the growing mode usually dominates the dynamics, but for k-gap solitons, we see exponential decay of the soliton in time. The nonlinearity creates an almost complete transfer of power from the growing modes to the decaying modes. As the intensity goes down, any slight variation in the initial conditions will inject energy to the growing mode, and the process will repeat indefinitely, resulting in an infinite train of solitons, which are equally spaced in time and do not interact with one another. This train of k-gap solitons emerges naturally under almost any initial conditions, extracting its energy from the modulation. The emergence of such a train of k-gap solitons is discussed in the Supplementary Material.

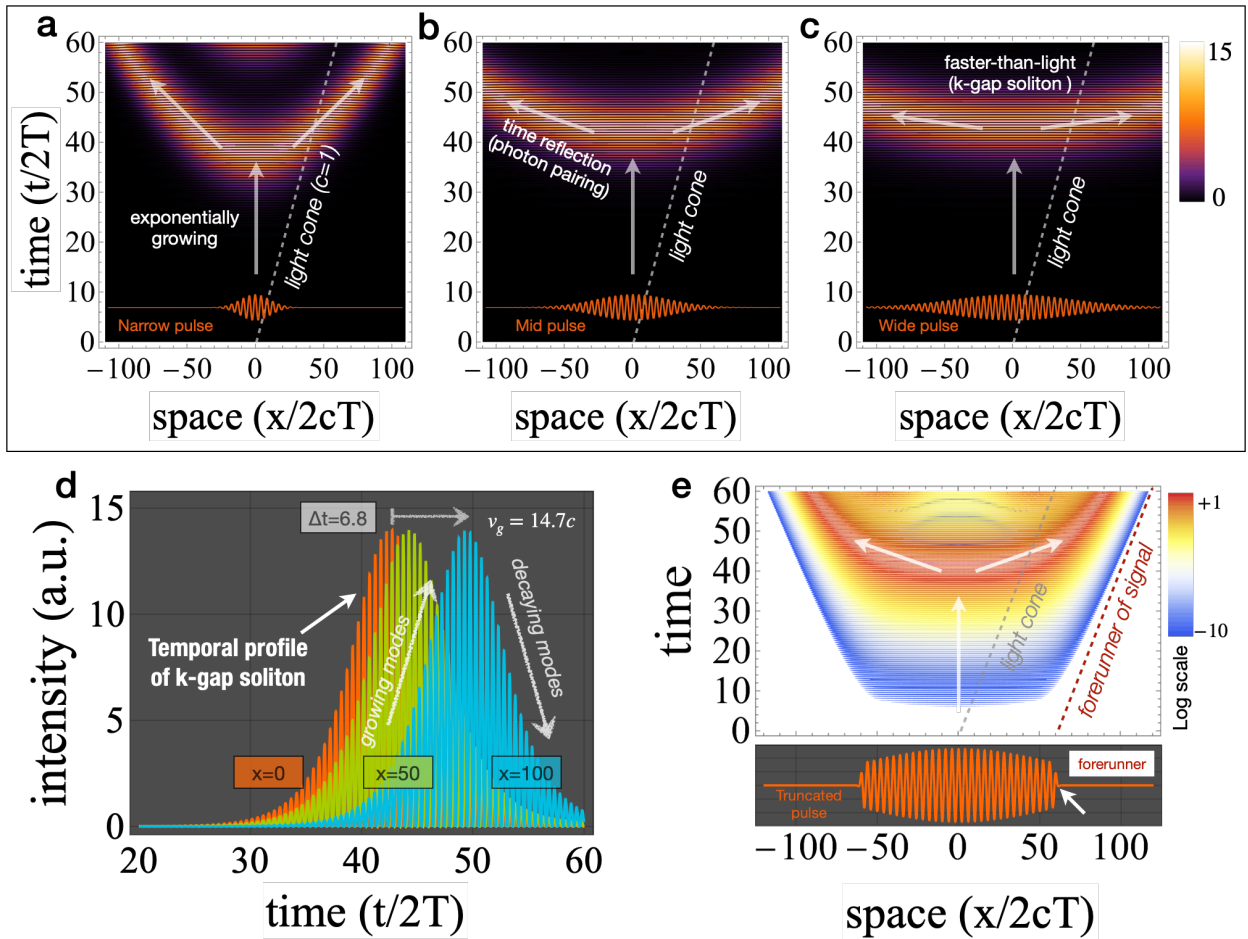

Fig. 2: (a-c) Generation of a k-gap soliton from an input of (a) narrow, (b) mid, and (c) wide beams, respectively. (d) The temporal profiles of the k-gap solitons at different locations ($x = 0, 50, 100$ [$2cT$]) found numerically from the seed beam in (b). This soliton exhibits superluminal behavior

with effective average group velocity of $v_g = 14.7c$. (e) Evolution of a truncated beam with a central wave vector ($k_0$) at the middle of the k-gap. The truncation mimics the forerunner of signal velocity, instead of the center-of-mass light cone. Once formed, the k-gap soliton travels faster than light, but still slower than the forerunner of the information in the leading edge.

Proceeding to the upper branch (+), similarly, we obtain the full k-gap soliton for a defocusing nonlinearity ($\gamma < 0$)

$$D(x,t) = \sqrt{2} u_0 \operatorname{sech}\left(\frac{t}{\tau_0}\right)\left(\cos\left((k_0 + \kappa - q)x - \frac{\Omega t}{2}\right) - \cos\left((k_0 + \kappa - q)x + \frac{\Omega t}{2}\right)\right) \quad (15)$$

where the frequency and wavevector conform with the 'linear' k-gap, $\omega = \frac{\Omega}{2}$, $k = k_0 - \kappa + q = k_0 - \kappa + \frac{3|\gamma|u_0^2}{4} \in k$-gap. The detailed derivations are given in the Supplementary Material.

**Generating k-gap solitons from a localized input**. The k-gap soliton is self-trapped in time but uniform and infinite in space. This raises a natural question on the ability to generate a k-gap soliton from a finite input beam. This issue is further highlighted by the superluminality of the k-gap soliton. To that end, we simulate the evolution of the k-gap soliton from a finite input "seed" beam with a limited bandwidth. We launch a weak Gaussian beam into the nonlinear PTC, with all its k-components located in the k-gap, and solve Eq. (3) numerically. The evolution of three input beams with spatial width ranging from narrow to wide are presented in Fig. 2(a-c). Initially, the intensity is weak, hence the input beam, being associated with the k-gap, exhibits exponential growth in time but no propagation dynamics in space ($v_g = 0$). Since the seed intensity is weak, the nonlinearity is still negligible, and the k-gap-induced amplification dominates the evolution. Since all the k-components are in the k-gap, they experience a fixed subharmonic drive frequency ($\Omega/2$). As time progresses, the growing field is strong enough and the nonlinearity comes into play, arresting the growth as the soliton reaches peak amplitude $u_0$. The arrest of the intensity growth happens because the band structure changes (through the nonlinear interaction) such that the wavepacket resides in the band rather than in the gap (See SM). Once the beam crosses into the band, the peak splits into two counterpropagating wavepackets [3]. Another feature we observe is how the propagation of the peak becomes superluminal. Taking closer look into the movement of the divided wavepacket, we note that, once reaching maximal intensity, the peak travels at superluminal velocity (Fig. 2). The field at different positions reaches the intensity apex at different times, and we can track the envelope trajectory and define an effective group velocity. We find that this field envelope preserves its temporal soliton-like shape with group velocity exceeding the speed of light in the medium (Fig. 2d and the supplementary videos in SM file). Figure 2d presents the temporal profiles at different positions for the simulation presented in Fig. 2b. We estimate that the superluminal apex propagates from $x = 0$ to $x = 200cT$ within $\Delta t = 13.6T$, so that the averaged group velocity over that distance is about $\bar{v}_g = 14.7c$. Comparing the formation of k-gap solitons for narrow (Fig. 2a), mid (Fig. 2b), and wide (Fig. 2c) excitation beams, we find that $\bar{v}_g$ of the peak is higher as the excitation beam is wider. The simulations indicate that the k-gap soliton will have infinite group velocity ($\bar{v}_g \to \infty$) in the limit of plane-wave (beam of infinite width) excitation, evolving into the theoretically predicted profile (Eq. 13). In these simulations, the parameters are $c = 1$, $\delta_1 = 0.12$, $\beta = 0.01$, $\Omega = 4\pi$, time is presented in units of 2T, and the space in the units of 2cT.

Next, we verify that this faster-than-light propagation of the k-gap soliton does not contradict Einstein's causality, by comparing the group velocity and the velocity of the information. Generally, the group velocity of a wavepacket is defined by the motion of its center, hence we launch a truncated Gaussian beam into the PTC, to have a clear cut on the propagation of the information contained in the leading edge. Figure 2e shows the dynamics of the truncated Gaussian seed. The truncated beam contains energy in band modes whose momentum components are outside the k-gap, and are therefore unamplified. The simulation shows that the truncation position (the forerunner) travels almost exactly at the speed of light in medium, but never above it. This issue is further highlighted by including any kind of material dispersion $\epsilon(\omega)$ and recalling that in the limit $\omega \to \infty$, the electromagnetic response is always $\epsilon \to 1$. The sharp edge (the forerunner) created by the truncation consists of all spatial wavenumbers, hence the forerunner always moves exactly at $c_0$, the speed of light in vacuum. As seen in Fig. 2e, all the momentum constituents of the k-gap soliton never go beyond the forerunner of the information. Further investigation shows that the velocity of the peak of the k-gap soliton is dependent on the seed's spatial profile. As the initial beam is wider, the apex of the soliton can move faster, and then slows down as it gets closer to the forerunner. Moreover, in principle, the soliton can arise from quantum fluctuations, as can be conjectured from recent work in quantum phenomena in PTCs [12,15]. In that case, the k-gap soliton would be created from arbitrary small noise by the amplification from the k-gap. However, even via the quantum process, the presence of the slightest signal cannot overrun the forerunner (see Fig. 2e).

**Power dependence of k-gap and threshold of amplification.** We find that the presence of the Kerr nonlinearity can alter the band structure of the PTC by shifting the k-gap to higher k-vectors, and shrinking the k-gap width. Under the mean-field approximation, given $n(t) = n_0 + n_1 \cos \Omega t + n_2 I$, with $n_0 = \sqrt{\epsilon_r}$, $n_1 = \chi_1/2$, $n_2 = \chi_3/2$, and $I = \frac{\epsilon_r}{2} \langle |E|^2 \rangle$ being the field intensity, one can solve the nonlinear energy band by linearizing around the k-gap (see the derivation in SM file), which yields

$$\omega(k,I) = \frac{\Omega}{2} \pm i\Omega \sqrt{\frac{1}{4}\left(\frac{n_1}{n_0+n_2 I}\right)^2 \left(\frac{c_0 k/\Omega}{n_0+n_2 I}\right)^4 - \left(\left(\frac{c_0 k/\Omega}{n_0+n_2 I}\right)^2 - \frac{1}{4}\right)^2} \qquad (16)$$

The field intensity ($I$) shifts the center of the k-gap. The k-gap amplification/gain factor is given by the imaginary part of (16): $\gamma(k,I) = |\Im(\omega(k,I))|$, that is, different k-components experience different gain factors. The arrest of the exponential growth occurs because, as the intensity increases, the shift in the band structure eventually brings the k-components of the soliton outside the momentum gap. A detailed explanation can be found in the Supplementary Material.

**Conclusion.** To summarize, we presented k-gap solitons in nonlinear PTCs, and found them to be superluminal. The faster-than-light behavior is understandable by Sommerfeld's forerunner, as tested numerically by the truncated seed beam. As a result, the superluminal k-gap soliton does not contradict Einstein's Special Relativity. The interplay between the time-modulation and the Kerr nonlinearity, mixes growing and decaying modes in the k-gap; it gives rise to the exponential growth in time until a certain peak is reached, and then surprisingly leads to the decaying intensity profile of the soliton in time, such that the shape of a k-gap soliton is always finite in its temporal width, although instability eventually leads to infinite train of solitons. Our results of superluminal k-gap solitons provide new insights into the study of time-varying media [51] and the gapped momentum states [52] in both photonics and other time-modulated physical systems.